\documentclass[aps,prl,twocolumn,superscriptaddress]{revtex4-2}

\usepackage{graphicx}
\usepackage{dcolumn}
\usepackage{bm}
\usepackage{color}
\usepackage{soul} 

\begin{document}

\title{Gelation of functional peptides by trivalent cations at the air-water interface}

\author{Stephen A. Crane}

\affiliation{Department of Chemical and Biomolecular Engineering, University of Pennsylvania, Philadelphia, PA, USA.}

\author{Felipe Jim\'enez-\'Angeles}

\affiliation{Department of Materials Science and Engineering, Northwestern University, Evanston, IL, USA.}

\author{Monica Olvera de la Cruz}
\affiliation{Department of Materials Science and Engineering, Northwestern University, Evanston, IL, USA.}

\author{Ivan J. Dmochowski}
\affiliation{Department of Chemistry, University of Pennsylvania, Philadelphia, PA, USA.}

\author{Kathleen J. Stebe}
\email{Contact author: kstebe@seas.upenn.edu}
\affiliation{Department of Chemical and Biomolecular Engineering, University of Pennsylvania, Philadelphia, PA, USA.}

\date{\today}

\begin{abstract}
We report a mechanism for gelation at fluid interfaces driven by multivalent-cation-mediated bridging. At the air-water interface, peptides with bound lanthanide cations undergo a coordination-geometry transition that converts the metal from a single-peptide bound state to a multi-peptide bridging state, driving charge inversion and gel formation. Surface adsorption and non-ideal interfacial electrostatics are implicated in this transition. The gel is stabilized by reversible metal-ligand coordination bonds that resist bulk salt screening, fundamentally distinct from electrostatic charge-inversion gelation in proteins. This reveals the breakdown of the peptide's coordinating sphere as a distinct pathway for interfacial gelation, independent of the diffuse electrostatic mechanisms governing bulk protein aggregation.

\end{abstract}

\maketitle

Multivalent ions in soft matter modify intermolecular forces and trigger unexpected events including polyelectrolyte precipitation \cite{widom1983monomolecular,raspaud1998precipitation}, redissolution \cite{solis2000flexible,de1995precipitation,hsiao2006salt}, and the adsorption of negatively charged molecules to negatively charged surfaces \cite{cheng2006polynucleotide}. According to Poisson-Boltzmann theory, charged macromolecules in solution are surrounded by a diffuse ionic structure that compensates for their charge. This picture breaks down in the presence of multivalent ions, which overcompensate the macromolecules' charge, leading to an effective charge inversion \cite{besteman2007charge}. For example, trivalent cations determine the adsorption of negatively charged proteins on solid surfaces, cross-link proteins at the air-water interface, and regulate the stability of foams \cite{fries2017multivalent,richert2019specific}. All of these phenomena have been linked to a reentrant condensation of the proteins in the bulk \cite{zhang2008reentrant}. Interactions with multivalent cations are often accompanied by a loss of structural integrity, especially for proteins or peptides at interfaces \cite{chen2011distinct,campbell2018adsorption,perriman2007effect}.   
Fluid interfaces introduce important additional complexities. For example, at interfaces, electrostatic and van der Waals forces can become comparable and give rise to specific ion effects \cite{dos2010surface,jungwirth2014beyond}. Furthermore, interfaces constrain molecular motion and create unique dielectric environments that modify molecular interactions \cite{schlaich2016water}, ion solubility \cite{levin2009ions}, and the kinetic properties of molecules.

\begin{figure}
\includegraphics[height=6.7cm]{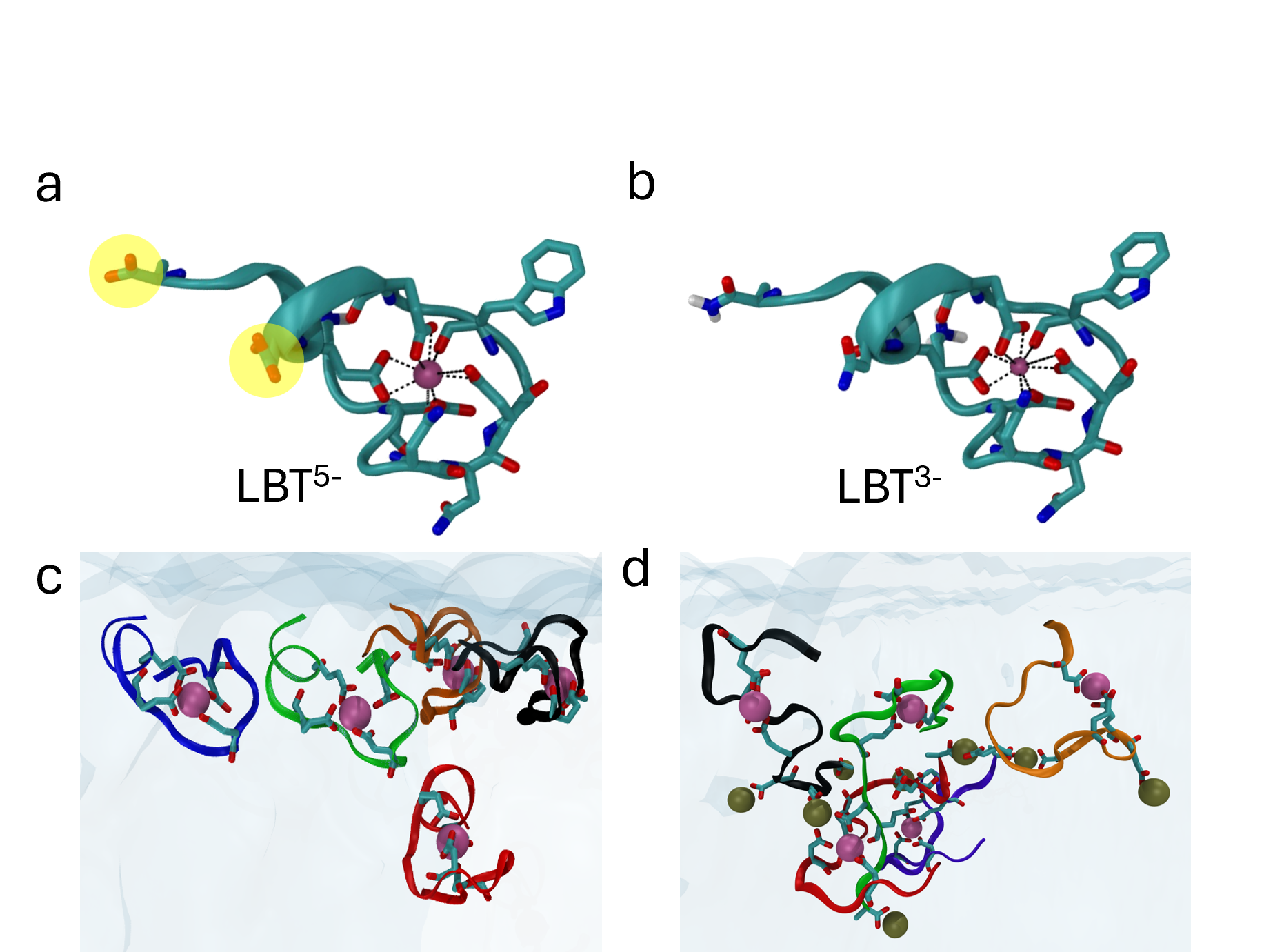}
\caption{\label{intro}  MD simulations of (a) LBT$^{5-}$ with bound Tb$^{3+}$.   Binding  between residues and cation is represented by dotted lines. Non-binding loop carboxylate ligands are highlighted. (b) Snapshot of the LBT$^{3-}$ Tb$^{3+}$  complex. (c) Intact binding loops and (d) compromised binding loops at the air-water interface.}
\end{figure}

Here, we study the interactions of  Tb$^{3+}$ cations with peptides adsorbed at air-water interfaces. The peptides, lanthanide binding tags (LBTs), selectively bind Tb$^{3+}$ in a carboxylate-rich binding loop with nanomolar affinity in bulk solution (Fig. \ref{intro}) \cite{franz2003lanthanide, nitz2004structural}. We consider two peptides; LBT$^{5-}$ (Fig. \ref{intro}a), with net charge $-5$ and LBT$^{3-}$ (Fig. \ref{intro}b), with net charge  $-3$, with the sequences:
\begin{center}
LBT$^{5-}$: NH$_3^+$-YI\textbf{D}T\textbf{N}N\textbf{D}G\textbf{W}Y\textbf{E}G\textit{D}\textbf{E}LLA-\textit{COO$^-$}
LBT$^{3-}$: NH$_3^+$-YI\textbf{D}T\textbf{N}N\textbf{D}G\textbf{W}Y\textbf{E}GN\textbf{E}LLA-CONH$_2$.
\end{center}
where the residues that coordinate the cation within the binding loop are shown in bold \cite{nitz2004structural,ortuno2024lanthanide}. In LBT$^{5-}$, the italicized \textit{D} and C-terminal \textit{COO$^-$} are anionic ligands that do not participate in selective cation binding; these ligands are modified via amidation in LBT$^{3-}$. In prior work, we have found that LBT$^{5-}$, but not LBT$^{3-}$, attracts excess cations to the air-water interface since the complexes are negatively charged, creating a mean electrostatic potential that favors cation adsorption. The excess cation adsorption leads to the formation of an interfacial gel \cite{crane2024interfacial}. 
In this study, we examine the structure of the peptide-cation complex at the interface and the physical properties of the gel in the presence of excess trivalent cations. We discover that, unlike globular proteins whose interfacial gelation follows complex denaturation pathways \cite{postel2003structure,gochev2024exploring}, gelation here is driven by coordination-shell reorganization. The adsorbed peptide-cation complex (Fig. 1c) loses its structural integrity (Fig. 1d), enabling highly persistent metal-mediated bridges between peptides. These bridges are stable even under bulk salt screening.

\begin{figure}
\centering
\includegraphics[height=15cm]{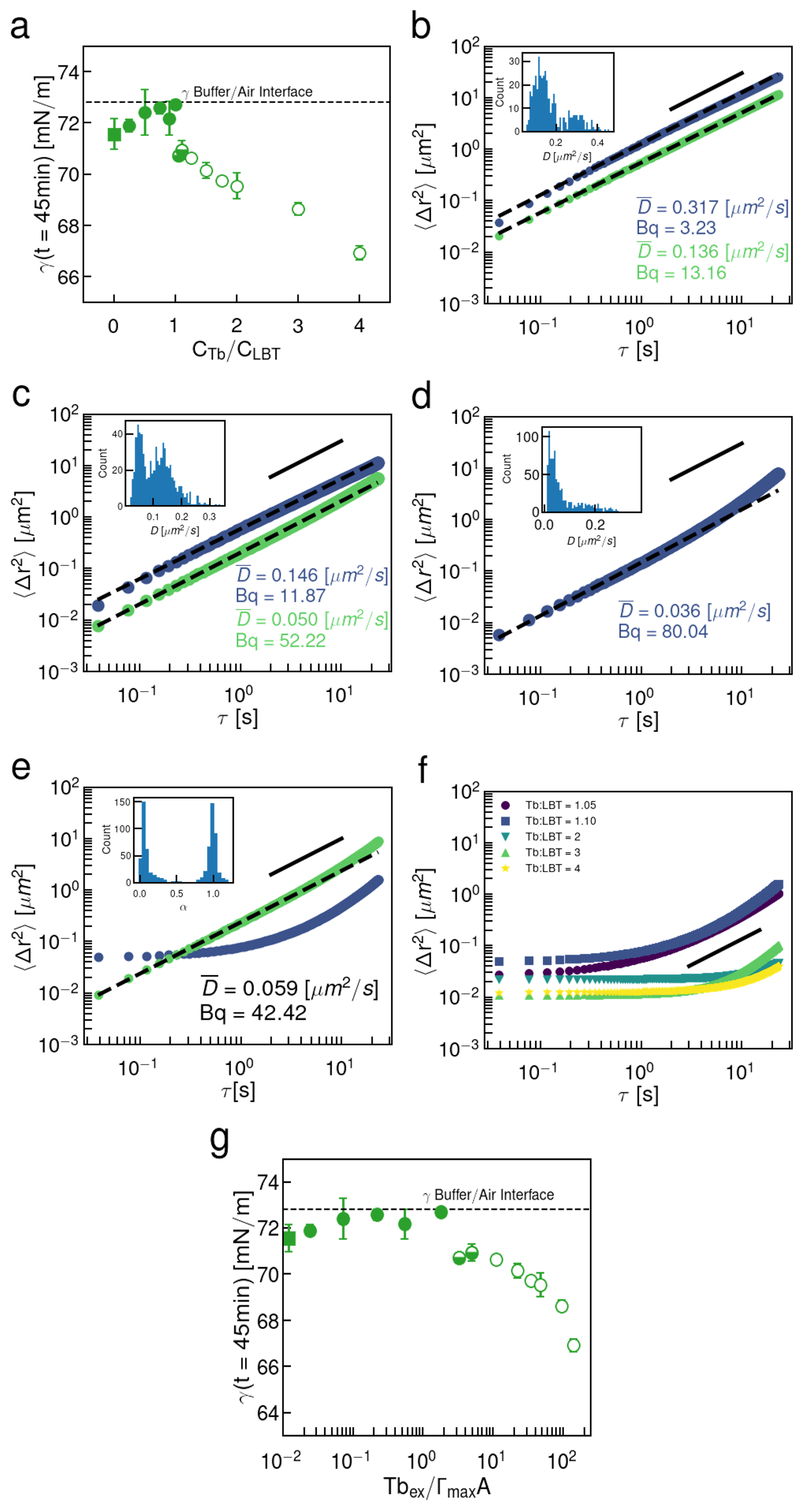}
\caption{\label{figlbt5msdst}(a) Apparent  surface tension of LBT$^{5-}$ solutions with varying ratios of Tb$^{3+}$ to LBT$^{5-}$. Closed symbols: viscous interface. Open symbols: gelled interface. Half-filled symbols: partially gelled interface. (b)-(d) MSD of tracer particles at the interface. Inset: distribution of particles' diffusivities. (b)  $C_{Tb}/C_{LBT}=0.5$. (c)  $C_{Tb}/C_{LBT}=0.75$ . (d) $C_{Tb}/C_{LBT}=1.0$ (e) MSD of tracer particles at interface. $C_{Tb}/C_{LBT}=1.1$. Inset: Distribution of power law exponents. (f)  MSD for $1.05 \leq C_{Tb}/C_{LBT}\leq4$. Solid blacklines: slope=1.0. Dashed black lines: power law fits to the ensemble MSD. (g) Surface tension versus  the ratio of Tb$^{3+}_{ex}$ to the maximum number of LBT$^{5-}$ that can be packed at the interface.}
\end{figure}

The apparent surface tension of an aqueous solution of  LBT$^{5-}$(Fig. \ref{figlbt5msdst}a), measured by the pendant drop method, changes non-monotonically with bulk cation concentration $C_{Tb}$. The peptide concentration is held fixed at  $C_{LBT}=65\ \mathrm{\mu M}$.  The surface tension increases in Regime I, for   $0\leq C_{Tb}/C_{LBT}\leq 1$. Thereafter, in Regime II, for $C_{Tb}/C_{LBT}$  slightly higher than 1.0, the apparent surface tension drops abruptly. Finally, in Regime III, for $C_{Tb}/C_{LBT}> 1.1$, the apparent surface tension monotonically decreases. 

To probe for interfacial gelation, we characterize the interfacial rheology  by measuring the mean-squared displacement (MSD) of tracer particles with radius $a=0.5\ \mu m$ dispersed on the interface. Figure \ref{figlbt5msdst} shows the ensemble-averaged MSD of the particles for $C_{Tb}/C_{LBT}$  from 0 to 4. In Regime I, shown in Fig. \ref{figlbt5msdst} b, c, \& d, the MSDs  follow a power law with exponent near unity, revealing a viscous interfacial layer. We find the surface viscosity by fitting the data to a power law  $MSD=4D\tau^\alpha$, where $\alpha$ $\approx$ 1.0. Given $D$, the Stokes-Einstein relationship and hydrodynamic theory are used to find the surface viscosity, $\eta_s$  \cite{hughes1981translational}, reported in terms of the Boussinesq number, $Bq=\eta_s/\eta a$, where $\eta$  is the shear viscosity of the  bulk solution.  The MSD data also reveal a bimodal distribution of tracer diffusivities (see Fig. \ref{figlbt5msdst}b \& c, and their insets); these data suggest that non-uniformly distributed peptide clusters form at the interface by cation bridging, with highly viscous regions being richer in clusters.  Clustering increases with cation concentration, slowing tracer motion. For $C_{Tb}/C_{LBT}$ of  0.5,  the Bq of the more and less mobile regions are 3.23 and 13.16, respectively, while for $C_{Tb}/C_{LBT}$ of  0.75, the Bq of the more and less mobile regions are 11.87 and 52.22, respectively. The surface viscosity increases everywhere, implying that the clustering increases continuously with cation concentration. Cluster formation would also reduce the number of  independent thermal  entities, causing surface tension to increase.  In Regime I, cation recruitment reduces electrostatic repulsion,  which also increases the surface tension.  For $C_{Tb}/C_{LBT}$ of  1.0 the fraction of more rapidly diffusing tracers is highly diminished, suggesting that almost all LBTs are clustered at the interface. 
In Regime II, for $C_{Tb}/C_{LBT}=1.1$ (Fig. \ref{figlbt5msdst}e), there are now two distinct populations of tracers. The inset reveals a bimodal distribution of the power law exponent near 1.0 and 0.0, indicating that some tracers sample  viscous regions, while others sample  a gel.  Regime III is probed for $2.0\leq C_{Tb}/C_{LBT}\leq 4.0$, (Fig. \ref{figlbt5msdst}f). The power law exponent near zero  indicates network formation. As cations are added in this regime,  the elastic regime lasts for longer lag times and the plateau shifts downward, implying a more robust interfacial network. 

We recast the apparent surface tension data against the ratio of Tb$^{3+}_{ex}$ divided by the peptide maximum packing $\Gamma_{max}$ (Fig. \ref{figlbt5msdst}g). Here,  Tb$^{3+}_{ex}$ is the total number of cations less those bound to peptides according to the dissociation constant characterizing peptide affinity, $K_D$, and  $\Gamma_{max}$  is the inverse  minimum cross sectional area per molecule at the interface determined previously via X-ray reflectivity  \cite{ortuno2024lanthanide}. In this form, we can more directly compare the experimental results to molecular dynamics (MD) simulation studies, described below. 

We perform MD simulations of LBT$^{5-}$ at the air-water interface using the GROMACS package \cite{pall2020heterogeneous,hess2008gromacs} and the CHARMM force field \cite{brooks2009charmm}. All simulations are initialized with 110 LBT$^{5-}$ peptides;  each peptide wraps a  Tb$^{3+}$ cation in its binding loop and is located at the air-water interface. The simulation box is 10 nm $\times$ 10 nm $\times$ 40 nm and also contains Na$^{+}$, Cl$^{-}$, and MES buffer molecules. Additional cations outside of the binding loop are added, defined as Tb$^{3+}_{ex}$. Water molecules are considered using the OPC model \cite{izadi2014building}. Specific details can be found in the Supplemental Materials (SM). 

\begin{figure}
\includegraphics[height=16cm]{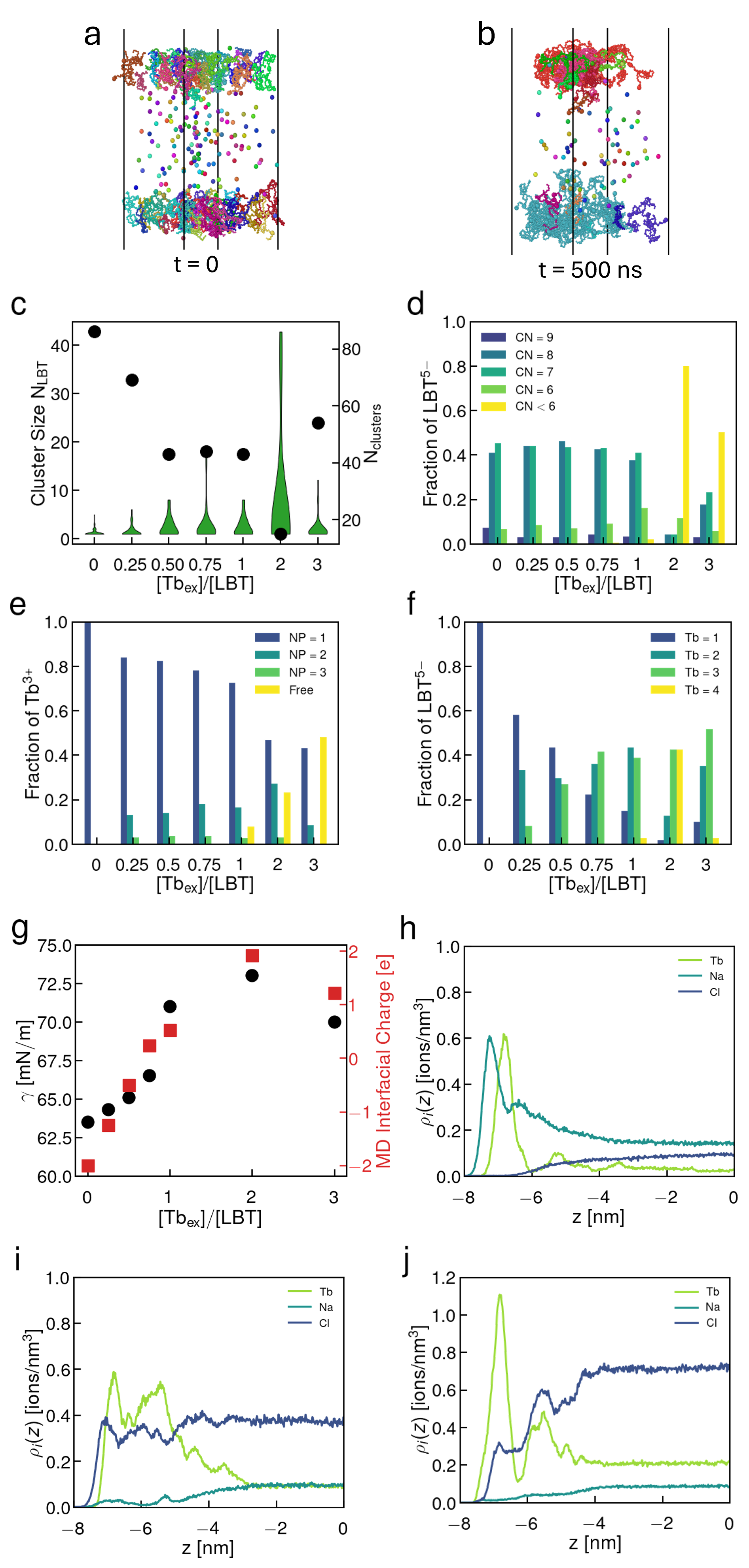}
\caption{\label{LBT5simulation} MD  simulations of Tb$^{3+}$ and LBT$^{5-}$ interactions. Snapshot of 2:1 ratio of  Tb$^{3+}_{ex}$  to  LBT$^{5-}$; independent clusters are indicated in distinct colors (a) Initially (at 0 ns), most LBT$^{5-}$ are independent. (b) After 500 ns,  LBTs are crosslinked. (c) Cluster size distribution (left axis) versus the ratio of Tb$^{3+}_{ex}$ to LBT$^{5-}$. Circles: Number of clusters for each ratio (Right axis). (d) Fraction of LBT$^{5-}$ at the interface with specified coordination number (CN) of the bound Tb$^{3+}$ in the binding loop versus the excess Tb$^{3+}$ per LBT$^{5-}$. (e) Fraction of Tb$^{3+}$ cations bound to different LBT$^{5-}$, where NP indicates the number of peptides = 1, 2, or 3 and those that are non-interacting (free) versus the excess Tb$^{3+}$ per LBT$^{5-}$. (f) Fraction of LBT$^{5-}$ peptides interacting with 1, 2, 3, or 4 Tb$^{3+}$ cations versus the excess Tb$^{3+}$ per LBT$^{5-}$. (g) Interfacial tension as a function of Tb$^{3+}_{ex}$ measured by MD simulation (circles). The calculated surface charge per peptide versus the excess Tb$^{3+}$ per LBT$^{5-}$ is shown on the second vertical axis (squares). (h) Ion density profile for Tb$^{3+}_{ex}/LBT^{5-}=0$. (i) Ion density profile for Tb$^{3+}_{ex}/LBT^{5-}=2.0$. (j) Ion density profile for Tb$^{3+}_{ex}/LBT^{5-}=3.0$.} 
\end{figure}

Simulations reveal crosslinked peptide clusters, as anticipated  by the interfacial rheology. Clustering is analyzed using a tool implemented in Ovito; details can be found in the SM \cite{ovito}. In Figure \ref{LBT5simulation}a,  independent entities, whether peptides or cation-peptide clusters, are shown in distinct colors. At  $t=0$ ns, each peptide is independent.  After 500 ns, as shown in Fig. \ref{LBT5simulation}b, most peptides are part of a few clusters.  The distribution of cluster sizes, shown in Figure \ref{LBT5simulation}c, increases with   $[Tb^{3+}_{ex}]/ [LBT^{5-}]$.  This trend continues for $[Tb^{3+}_{ex}]/ [LBT^{5-}]$ up to 2.0; the number of peptides in a cluster increases, and the number of independent clusters decreases. Cluster formation relies on reducing direct coordination of the selectively bound cations and the binding ligands, as quantified by the coordination number (CN) (see Fig. \ref{LBT5simulation}d).
  
For  dilute Tb$^{3+}_{ex}$ solutions, with concentrations below the peptide concentration, the preponderant CN is  7 or 8, indicating strong coordination of the binding loop cation. As  Tb$^{3+}_{ex}$ concentration increases, however, a significant fraction of binding loop cations have CN of 6 and below, indicating loss of binding loop integrity.  The binding loop opens, freeing its carboxylates to interact with the excess cations. Thereafter, the trend reverses; the number of clusters increases and their size decreases. For  $[Tb^{3+}_{ex}]/ [LBT^{5-}]$  of 3.0, the fraction of peptides with  CN $<$ 6 decreases, implying that the binding loops are less opened, and that the equilibrium shifts closer to the tightly bound conformation.  Since these loops expose fewer anionic ligands for crosslinking, the number of clusters increases, while the size of the clusters decreases.

The distribution of cations that interact with one or multiple peptides (NP) is reported in Figure \ref{LBT5simulation}e.  Absent excess cations, a single binding loop cation interacts with each peptide, and NP=1 for all cations.  As Tb$^{3+}_{ex}$ concentration increases, for $[Tb^{3+}_{ex}]/ [LBT^{5-}] < 1$,  an increasing number of cations interact with two or three peptides.  For $[Tb^{3+}_{ex}]/ [LBT^{5-}] \geq 1$, a population of free cations in solution  develops. Non-selective binding and network formation also increase the fraction of peptides that interact with with 1, 2, 3, or 4 cations, as shown in Fig. \ref{LBT5simulation}f; these cations include those shared between peptides and those that are bound by a single peptide. Notably, the fraction of peptides  interacting with three or more cations exceeds the fraction of cations that interact with three or more peptides, implying that the disordered peptides are the nodes of the network. 

The surface tension of the peptide-laden interface is determined from simulation using the diagonal components of the stress tensor \cite{alejandre1995molecular}, as shown in Fig. \ref{LBT5simulation}g, and is found to be non-monotonic with cation concentration. For   $[Tb^{3+}_{ex}]/ [LBT^{5-}]$ up to 2.0, the surface tension increases.  For concentration ratios greater than 2.0, the surface tension decreases. These trends reflect the effects of surface charge and cluster formation.  In the first regime, excess cation recruitment leads to charge reversal at the interface \cite{jimenez2008regimes}.
We calculate the mean charge per adsorbed peptide  $Q$:  $Q=3\mathrm{Tb^{3+}_{int}/LBT^{5-}}-5,$ where $\mathrm{Tb^{3+}_{int}=(Tb_{ex}/LBT+1)(1-Tb_{Free}/Tb_{total})}$. 
We find that the surface charge per peptide rapidly approaches 0.0 from $-$2.0 as slight amounts of Tb$^{3+}_{ex}$ are added.  The initial increase in surface tension can be attributed to two factors:  the decrease in the number of independent thermal entities  as clusters form, and the decrease in electrostatic repulsion.  The maximum surface tension occurs for $[Tb^{3+}_{ex}]/ [LBT^{5-}]$ of 2.0 at the interface.  Here, the interfacial charge per peptide has already reversed and has a value of 1.9$e$, but the number of independent entities is minimized, so repulsion among them is not dominant. Finally, for $[Tb^{3+}_{ex}]/ [LBT^{5-}]$ of 3.0, the surface tension decreases. The number of clusters at the interface has increased by nearly a factor of three times those present for the  $[Tb^{3+}_{ex}]/ [LBT^{5-}]$ of  2.0 case, and the interfacial charge is still positive, $\sim$1.2$e$ per peptide. The many clusters interact and repel,  allowing  the tension to decrease.  

It is remarkable that, even  after the interfacial charge reversal, Tb$^{3+}$ cations continue to adsorb to the monolayer. The compensation of interfacial charge is evident in the ion density profiles perpendicular to the interface. For $[Tb^{3+}_{ex}]/ [LBT^{5-}]<0.25$ the net-negative surface charge is compensated by free Na$^+$ ions in solution. As predicted by Poisson-Boltzmann theory, the Na$^+$ ions exhibit an exponentially decaying profile, while that of  Cl$^-$ concentration increases, with a screening Debye length of $\lambda_D\approx0.7-0.8$ nm as shown in Fig. \ref{LBT5simulation}h. As excess Tb$^{3+}$ increases, the  trivalent cations adsorb and the Na$^{+}$ ions are released from the interface, similar to the liberation of counter ions during  adsorption of a polyelectrolyte approaching an oppositely charged surface \cite{sens2000counterion}. This adsorption is apparent in the ionic profiles for $[Tb^{3+}_{ex}]/ [LBT^{5-}] = 2.0$ and $[Tb^{3+}_{ex}]/ [LBT^{5-}] = 3.0$ in Fig. \ref{LBT5simulation}i \& j, which exhibit oscillatory behavior that does not conform to the predictions of Poisson-Boltzmann theory. 

Interestingly, when adsorbed at the interface, the Tb$^{3+}$ ions remain fixed, whereas the Na$^{+}$ and Cl$^{-}$ ions are mobile. The adsorbed Tb$^{3+}$ ions overcompensate the interfacial charge, reversing the sign of the electric field, and driving formation of the Cl$^{-}$ screening cloud. Such charge adsorption decreases the peptides' electrostatic energy, even if their charge is overcompensated \cite{solis2000flexible,10.1063/1.480763}. The electrostatic interaction between the ions and the charged groups is given as $U_{el}(r)=\frac{Q_iQ_j}{4\pi \epsilon_0 \epsilon_r r}$, where $Q_i$ and $Q_j$ represent the ion and residue charge, $r$ is their separation distance, $\epsilon_0$ is the vacuum permittivity, and $\epsilon_r$ is the effective dielectric constant due to the water molecules. Near the air-water interface, the water dielectric constant is $\sim$10 or less, while in bulk it is $\sim$80 \cite{schlaich2016water}. Thus, electrostatic interactions, especially between a trivalent cation and carboxylate anions, are made even stronger ($>k_BT$), driving excess cation recruitment. The strong interactions between excess Tb$^{3+}$ ions and interfacial acidic groups lead to the tearing apart of the binding pocket. As a result, each peptide has multiple binding sites that can connect with other peptides through Tb$^{3+}$ ions to form a network. Thus, the formation of a gel is a signature of these processes.

In this scenario, repulsion from the interface is long-ranged but relatively weak due to screening from the background electrolyte. Thus, as Tb$^{3+}$ cations entropically explore the solution, cations can approach  dangling carboxylates from LBT$^{5-}$ and be strongly attracted to the interface, and remain there once recruited. This is how nearly all of the LBT$^{5-}$ peptides at the interface link together through excess Tb$^{3+}$ as shown in Fig. \ref{LBT5simulation}b for $[Tb^{3+}_{ex}]/ [LBT^{5-}]$ of  2.0, and the reason that cations continue to be recruited at the interface for  $[Tb^{3+}_{ex}]/ [LBT^{5-}]$ of  3.0.  This latter regime's large number of  repulsive clusters is reminiscent of re-entrant phase behavior seen in protein solutions \cite{richert2019specific}.

With the insights gained from  simulation and interfacial rheology, we can explain the experimental surface tension shown in Fig. \ref{figlbt5msdst}g.  In Regimes I and II, the surface tension data trends in experiment and simulation  qualitatively agree. In Regime I, the experimental surface tension rises as the interfacial layer initially reduces its charge, crosslinks and reduces the number of  independent thermal entities. Rheology also suggests such clustering in the bimodal diffusivities of tracer particles. The maximum experimental surface tension occurs around 1.8 Tb$^{3+}_{ex}$ per LBT at the interface. Under these conditions, simulation shows the largest clusters in an overcharged interface, and rheology shows coexisting gelled and viscous domains at the interface, i.e. some gel clusters are large enough to trap the tracer particles, while others are too small and make a viscous suspension in the interface. 

In Regime III, the surface tension monotonically decreases in both simulation and experiment. However, the underlying mechanisms differ. In simulation, the number of repulsive  independent entities increases. However, unlike simulation,  the interface in experiment is in contact with a bulk source of peptides and cations that form a gel network at the interface. Rheology clearly shows that the interface becomes increasingly crosslinked.  We have two hypotheses for the ensuing apparent surface tension reduction. First, the recruitment of excess Tb$^{3+}$ can generate repulsion within the gel.  Additionally, once the gel covers the interface, any strains in the layer could reduce the apparent surface tension. 

\begin{figure}
\includegraphics[height=10cm]{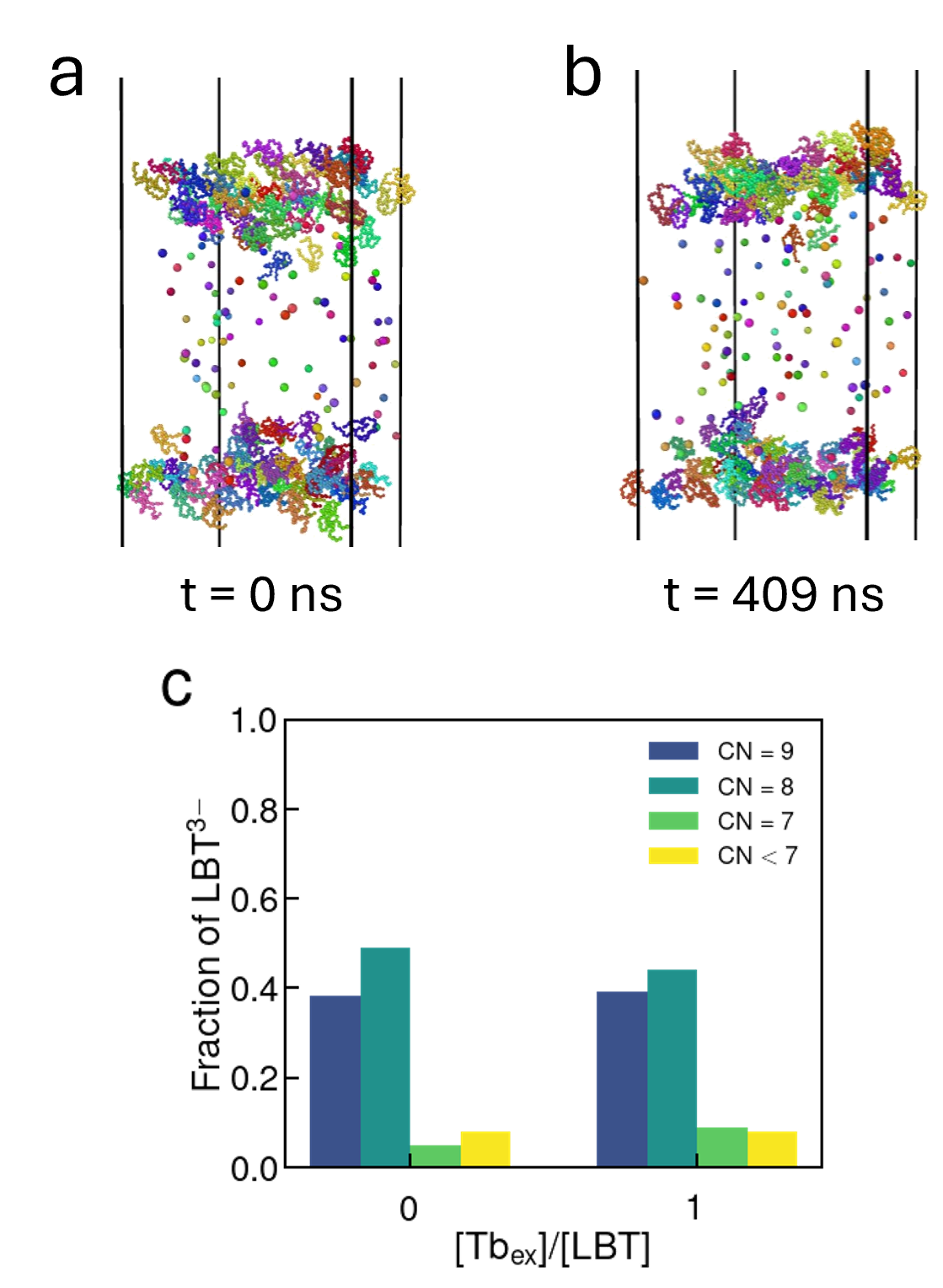}
\caption{\label{lbt3} MD  simulations of Tb$^{3+}$ and LBT$^{3-}$ interactions. Snapshot of 1:1 ratio of  Tb$^{3+}_{ex}$  to  LBT$^{5-}$; independent clusters are indicated in distinct colors (a) Initially (at 0 ns), all LBT$^{3-}$ are independent. (b) After 500 ns,  LBT$^{3-}$ remain independent. (c) Fraction of LBT$^{3-}$ at the interface with specified coordination number (CN) of the bound Tb$^{3+}$ in the binding loop versus the excess Tb$^{3+}$ per LBT$^{3-}$.}
\end{figure}

The destruction of the binding loop in the presence of excess cations is not inevitable. The peptide  LBT$^{3-}$ forms net neutral complexes without exposed anionic ligands. For solutions of this peptide, the surface tension decreases and the interfacial  viscosity increases as Tb$^{3+}$ is added, suggesting that adsorption of peptide-cation complexes increases with cation concentration. There is no evidence of gelation even in the presence of significant excess Tb$^{3+}$ (see SM).  MD simulation with 1.0 Tb$^{3+}_{ex}$ per LBT$^{3-}$, shown in  Figure \ref{lbt3}a \& b at $t=0$ ns and $t=409$ ns, respectively, shows that excess cations are free in solution and  each peptide remains independent. The coordination numbers for Tb$^{3+}$ bound to the peptides are  predominantly at 8 or 9, Fig. \ref{lbt3}c. These results imply that the binding loop is well behaved at the interface.  We have studied this structure in prior work using X-ray reflectivity to quantify peptide adsorption and X-ray fluorescence near total reflection to quantify the number of cations at the interface, and find that cations and peptides are indeed present in a nearly 1:1 ratio \cite{ortuno2024lanthanide}. Neutrality of the peptide-cation complex does not suffice to protect against attack by excess cations;  the LBT peptide RR-LBT$^{5-}$, with exposed non-binding anionic groups undergoes interfacial gelation in spite of forming a neutral complex  \cite{crane2024interfacial}.

The LBT peptides have generated tremendous interest for their ability to  selectively capture REEs; we are interested in using this ability to capture REEs in foams. In bulk solution, the binding affinity of both LBT$^{5-}$ and LBT$^{3-}$ varies among the REEs in a similar manner.  Selective capture in foams relies on the  retention of  selective binding at the interface. Guided by the results in this Letter, we have recently demonstrated that LBT$^{3-}$ can indeed selectively capture Dy$^{3+}$ over Nd$^{3+}$ at air-water interfaces \cite{crane2026acridone}, supporting a green foam-based separation and recovery of elements with urgent societal significance.

To conclude, we study two peptides that are rich in anionic moieties.  The peptides selectively coordinate REE cations by multi-dentate coordination with carbonyl and carboxy- groups in a binding loop structure. The fate of this binding loop at the fluid interface depends on the presence of exposed anionic ligands outside of this loop, which  attract excess multivalent cations to the interface. We show that excess cations catastrophically disrupt the binding loop structure, leading to gelation, and, at high enough cation concentration, charge reversal. The reduced dielectric constant in the vicinity of the interface and the resulting strengthening of electrostatic interactions between multivalent cations and exposed anionic species are implicated in these gelation processes. Elimination of exposed ligands changes the scenario; LBT$^{3-}$ neither attracts excess cations, nor loses its binding loop structure at the interface. These findings suggest a criterion for the preservation of functional peptides at interfaces. Charged ligands that attract multivalent ions must not be exposed.

\begin{acknowledgments}
\textit{Acknowledgments$-$}This research was supported by the Department of Energy DOE-BES Award DE-SC0022240 and by the U.S. National Science Foundation under award number 1943384. We thank the Vagelos Insitute for Energy Science and Technology Graduate Fellowship for their support of SAC.

\textit{Author contributions$-$}SAC and FJA contributed equally to this work. KJS, IJD, FJA, MOdlC designed and supervised the research and aided in the interpretation of results. SAC performed all experiments, analysis and interpretation. FJA performed and analyzed simulations. SAC and FJA drafted the manuscript, which was reviewed and edited by all.
\end{acknowledgments}

\bibliography{bibliography}

\end{document}